\documentclass[12pt,preprint,twocolumn]{aastex63}
\usepackage{graphicx}   
\usepackage{amsmath}

\def \fps@figure{htbp}
\makeatother
\def \aap  {A\&A}

\def \apjs  {ApJS}

\def \apjl {ApJL}

\newcommand{\TNG}{IllustrisTNG}

\DeclareGraphicsExtensions{.ps,.pdf,.png}
\makeatletter

\shorttitle{Barred galaxies in TNG100}
\shortauthors{Zhao et al.}

\begin{document}

\title[TNG100 Bars]{Barred Galaxies in the \TNG\ Simulation}

\correspondingauthor{Min Du}
\email{dumin@pku.edu.cn}

\author{Dongyao Zhao}
\affiliation{Beijing Planetarium, Beijing Academy of Science and Technology, Beijing, 100044, China}
\affiliation{The Kavli Institute for Astronomy and Astrophysics, Peking University, 5 Yiheyuan Road, Haidian District, Beijing, 100871, China}

\author{Min Du}
\affiliation{The Kavli Institute for Astronomy and Astrophysics, Peking University, 5 Yiheyuan Road, Haidian District, Beijing, 100871, China}

\author{Luis C. Ho}
\affiliation{The Kavli Institute for Astronomy and Astrophysics, Peking University, 5 Yiheyuan Road, Haidian District, Beijing, 100871, China}
\affiliation{Department of Astronomy, Peking University, 5 Yiheyuan Road, Haidian District, Beijing, 100871, China}

\author{Victor P. Debattista}
\affiliation{Jeremiah Horrocks Institute, University of Central Lancashire, Preston, PR1 2HE, UK}

\author{Jingjing Shi}
\affiliation{The Kavli Institute for Astronomy and Astrophysics, Peking University, 5 Yiheyuan Road, Haidian District, Beijing, 100871, China}
\affiliation{Kavli IPMU (WPI), UTIAS, The University of Tokyo, Kashiwa, Chiba 277-8583, Japan}

\begin{abstract}
Almost two-thirds of disk galaxies in the local universe host bars, which serve as important drivers of secular evolutionary processes. While cosmological simulations are powerful tools to study the formation and evolution of galaxies, they have often struggled to generate reasonable bar populations. We measure the fraction, size, and strength of bars in 3866 disk galaxies from the TNG100 run of the advanced cosmological simulation \TNG.  Consistent with observations, about 55\% of disk galaxies with stellar mass $M_* \approx 10^{10.6}\, M_\odot$ are barred, and the relation between bar size and total stellar mass is similar to that found in near-infrared surveys. However, the formation of bars is suppressed in galaxies with $M_* < 10^{10.6}\, M_\odot$, which may result from the difficulty TNG100 has in resolving short bars with radius $<1.4$ kpc. In contrast, up to 75\% of massive disk galaxies with $M_* > 10^{10.6}\, M_\odot$ have bars, $\sim 10\%-20$\% higher than observed. TNG100 overproduces relatively short bars (radius $\sim 1.4-3$ kpc) with respect to the mass-bar size relation observed in near-infrared surveys. Tracing the progenitors of $z=0$ massive galaxies we find that the bar fraction increases from 25\% to 63\% between $z=1$ and $0$. Instead if we select all disk galaxies during $z=0-1$ with a constant mass cut of $M_* \geqslant 10^{10.6}\, M_\odot$ we find that the bar fraction is a nearly constant 60\%.\\
\\
\textit{Unified Astronomy Thesaurus concepts:} SB Galaxies (136); Galactic structure (622); Astronomy simulations (1855)\\
\end{abstract}

\section{Introduction} 
\label{sec:introduce}

Observational surveys show that about two-thirds of disk galaxies in the local Universe host a bar \citep[e.g.,][]{Eskridge00, MDelmestre07, Erwin18}. Galactic bars are expected to play an important role in the secular evolution of disk galaxies (see the reviews by \citealt[]{KormendyKennicutt04, Kormendy2013}). Bars can funnel gas efficiently toward the central regions of galaxies \citep[e.g.,][]{Athanassoula92,KimWT2012,LiZhi15}, possibly triggering nuclear starbursts \citep{HuntMalkan99, Jogee05} and even fueling active galactic nuclei \citep[e.g.,][]{Ho97, Lee12, Cheung13, Goulding17}. In turn, short bars of $\sim 1$ kpc radius can be destroyed by the growth of central concentration of galaxies \citep{Du2017}. The destruction of such bars may contribute to the growth of bulges without a  merger involved \citep{GuoMH2020}. Furthermore, the vertical buckling instability of bars produces boxy/peanut-shaped bulges \citep{Raha1991, Merritt&Sellwood1994}, which has considerable observational support in our own Galaxy \citep{Shen2010} and other galaxies \citep{Erwin&Debattista2016, LiZY2017}.

Bars form quickly once a dynamically cool disk has settled \citep[see the review by][and references therein]{Sellwood2014}. They are likely to grow longer and stronger by transferring angular momentum outward into the dark matter halo \citep{Debattista&Sellwood1998, Debattista&Sellwood2000, Athanassoula03}. The internal dynamics of how and why bars form has been addressed in many reviews \citep{Toomre1981, SellwoodWilkinson93, Binney&Tremaine2008, Sellwood2013}. However, it is still unclear why some galaxies have bars, while others do not \citep{Sellwood2019}. In order to better understand how bars form and what role bars really play in the evolution of disk galaxies, it is fundamental to determine what kind of galaxies do or do not host bars. 

The bar fraction has long been studied observationally. Early photographic work \citep[e.g., ][]{deVaucouleurs91} found that 65\% of bright nearby galaxies host bars. The bar fraction decreases to $\sim 30\%$ if only strong bars are considered \citep[see also][]{Sandage87}. These results have been supported by subsequent optical and near-infrared (NIR) studies. For example, NIR observations show that about $60$\% of local disk galaxies host bars \citep{Eskridge00, Knapen00, MarinovaJogee07, MDelmestre07, DiazGarcia16, Erwin18}, whereas studies based on the Sloan Digital Sky Survey \citep[SDSS;][]{York00} find a significantly lower total bar fraction ($25\%-40\%$) \citep{Barazza08, Aguerri09, NairAbraham10a, NairAbraham10b, Masters11, Masters12, Skibba12, Gavazzi15, Consolandi16}. This disagreement is likely due to the limitations of SDSS data \citep{Erwin18}. The moderate spatial resolution of SDSS images may preclude the identification of short and weak bars that are common in less massive galaxies. 

The incidence of bars depends on the stellar mass of the host galaxy. NIR studies \citep[e.g.][]{DiazGarcia16, Erwin18} show that the bar fraction increases with stellar mass in less massive galaxies with $M_* < 10^{9.7}\, M_\odot$, but remains nearly constant at $50\%-60\%$ in more massive galaxies. Bar sizes also depend on galaxy stellar masses. \citet{DiazGarcia16} and \citet{Erwin18, Erwin19}, using NIR data from the Spitzer Survey of Stellar Structure in Galaxies \citep[S$^4$G;][]{Sheth2010}, find a bimodal relationship between bar size and stellar mass: bar size is nearly constant at $\sim 1.5$ kpc in galaxies with $M_*\lesssim 10^{10.2}\, M_\odot$, while at higher $M_*$ the bar size scales as $\propto M_*^{0.56}$. 

Numerical simulations are powerful tools to study the formation and evolution of bars, and the role they play in secular evolution. Although the dynamical influence of bars has been studied in great detail using individual cases \citep[e.g.,][]{Curir06, Kraljic12, Guedes13, Goz15, Debattista19}, a full, systematic understanding of bars needs a statistically large enough sample of barred galaxies from cosmological simulations. Recent cosmological simulations have been able to generate realistic galaxies with reasonable bulge-to-disk ratios \citep{Huertas-Company2019, Park2019, Tacchella2019, Du2020}, as a result of the significant progress made in modelling galaxy formation physics \citep{Agertz2011, Guedes2011, Aumer2013, Stinson2013, Marinacci2014, Roskar2014, Murante2015, Colin2016, Grand2017}. Additionally, increasing computational power has permitted an increased resolution of such simulations, enabling bars to be resolved above a certain mass limit. However, it is still challenging to reproduce bar fractions as high as those measured in observations. For example, \citet{Algorry17} showed that only $\sim 40\%$ of massive disk galaxies with $M_* = 10^{10.6} - 10^{11}\, M_\odot$ in the EAGLE simulation \citep{Crain2015, Schaye2015} have bars at $z=0$. The bar fraction in the original Illustris cosmological simulation \citep{Genel14,Vogelsberger14} is even lower (26\%) for disk galaxies with $M_* > 10^{10.9}\, M_\odot$ \citep{Peschken19}. 

The advanced version of Illustris, named \TNG, reproduces realistic galaxies that successfully emulate real galaxies in many aspects. Recent studies have concluded that the bar fractions amongst disk galaxies in the \TNG\ simulation are consistent with observations. For example, \citet{Yetli20} found that bars can be detected in 40\% of \TNG\ disk galaxies with $M_* > 10^{10.4} M_\odot$ at $z=0$, and \citet{Zhou20} reported that 55\% of disk galaxies with $M_* > 10^{10.5} M_\odot$ have bars. However, these studies use the Fourier method, while bars are generally identified by an ellipse fitting method in observations. In this paper, we revisit the bar fraction of disk galaxies in the \TNG\ simulation by the ellipse fitting method, and we compare systematically our derived bar sizes with observations. 

This paper is organized as follows. We introduce the \TNG\ simulation in Section~\ref{sec:TNG_simulation}. Section~\ref{sec:parent_disk} presents how the parent disk galaxies are selected. The methods we use to identify and measure bars are described in Section~\ref{sec:bar_inden}, where we also present a catalog of barred galaxies. Section~\ref{sec:result_z0} discusses the main results for redshift $z=0$. The evolution of the bar fraction at $z=0-1$ is presented in Section~\ref{sec:result_zgt0}. We summarize our main conclusions in Section~\ref{sec:conclude}.

\section{The \TNG\ Simulation}
\label{sec:TNG_simulation}

\TNG\ is an advanced magneto-hydrodynamical cosmological simulation \citep{Marinacci18, Naiman18, Springel18, Nelson18, Pillepich18_475} run with the moving-mesh code AREPO \citep{Springel10}. As described in \citet{Weinberger17} and \citet{Pillepich18_473}, \TNG\ uses an updated version of the Illustris galaxy formation model. We use the TNG100 run, the highest resolution version that is publicly accessible \citep{Nelson19}. It simulates a volume with side length $75 h^{-1} \approx 111$ Mpc. The average mass of the baryonic resolution elements in TNG100 reaches $1.39 \times 10^6\, M_\odot$. The gravitational softening length of the stellar particles is set to $0.5 h^{-1} \approx 0.74$ kpc. 

The optical morphologies of the galaxies in TNG100 are in good agreement with observations of local galaxies \citep{Huertas-Company2019, RodriguezGomez19}. A systematic comparison between the \TNG\ simulation and the Pan-STARRS survey showed that the optical size and shape of TNG100 galaxies are consistent with observations within the $\sim 1\,\sigma$ scatter of the observed trends \citep{RodriguezGomez19}. Beyond the basic morphological bulge-disk decomposition, TNG100 also successfully reproduces complex dynamic structures. For example, \citet{Xu19} found that the relative fractions of the cold, warm, and hot orbital components in TNG100 galaxies are remarkably consistent with those estimated from integral-field spectroscopic observations of nearby galaxies \citep{Zhu2018b}.  Furthermore, decomposing the simulated TNG100 galaxies into their intrinsic kinematic structures  \citep{Du2019} produces individual components that closely resemble physically familiar galaxy components \citep{Du2020}. 

In this paper, the dark matter halos in each simulation snapshot are from the catalog provided by the IllustrisTNG simulation, which are identified by using the friend-of-friend algorithm \citep{Davis85}, and subhalos in each halo are further identified using the {\tt SUBFIND} algorithm \citep{Springel01, Dolag09}. A galaxy is defined to be a gravitationally bound object with non-zero stellar mass within the halo or subhalo. We define the center of each galaxy as the minimum of the gravitational potential. The $z$-axis of each galaxy is oriented with the total angular momentum vector of its stars within 0.1 times the virial radius.

\section{Parent Sample of Disk Galaxies}
\label{sec:parent_disk}

We select 6507 galaxies at redshift $z=0$ with stellar masses $M_* \geqslant 10^{10.0}\, M_\odot$ calculated within a radius of 30 kpc. This criterion ensures that every galaxy has enough stellar particles ($>10^4$) to resolve its structure.  The sample of disk galaxies is then chosen based on the fraction of kinetic energy in ordered rotation \citep{Sales2010}. The parameter $k_{\rm rot}$ measures the mass-weighted average value of $v_\phi^2/v^2$ within 30 kpc, where $v_\phi$ is the azimuthal velocity and $v$ the total velocity of each star particle. Thus, $k_{\rm rot} = 1/3$ for a spheroidal galaxy that is completely dominated by random motions. Defining disk galaxies with the criterion $k_{\rm rot} \geqslant 0.5$ yields a parent sample of 3866 galaxies, or $59$\% of all galaxies in this mass range. This fraction is consistent with the observational frequency of disk galaxies of $M_* \geqslant 10^{10.0}\, M_\odot$ determined by \citet{Conselice06} using the Third Reference Catalogue of Bright Galaxies \citep{deVaucouleurs91}.

\section{Bar Detection and Characterization}
\label{sec:bar_inden}

\subsection{Method: Ellipse Fitting}
\label{sec:ellipse_fitting}

\begin{figure*}
\raggedright
\includegraphics[scale=0.208]{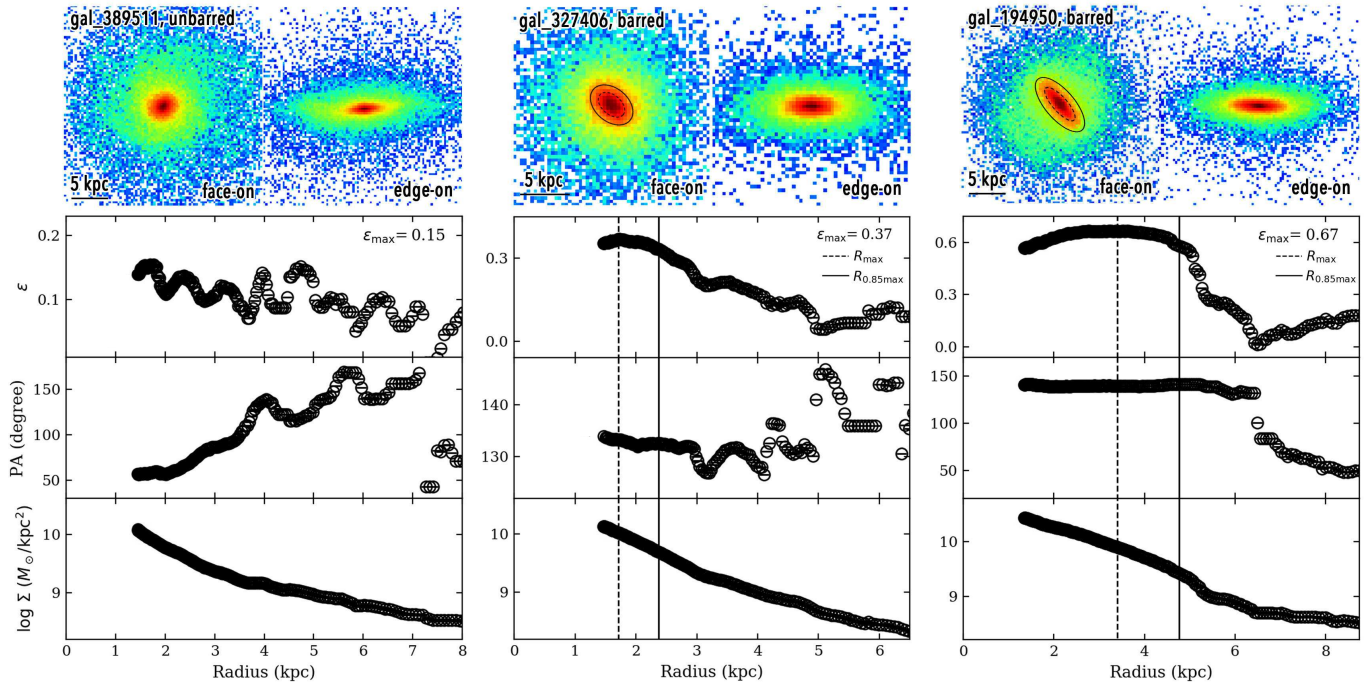}  
\caption{Example of (left) an unbarred galaxy, (middle) a weakly barred galaxy, and (right) a strongly barred galaxy identified by analysis of the elliptical isodensity contours of face-on mass surface density maps of TNG100 galaxies. The face-on and edge-on surface density maps are shown in the first row. The radial profiles of $\varepsilon$, PA, and surface density measured by {\tt ellipse} fitting are shown in the following rows. The maximum ellipticity ($\varepsilon_{\rm max}$) is given in the $\varepsilon$ panel. For the barred galaxies, the dashed ellipse and vertical dashed line mark $R_{\rm max}$, and the solid ellipse and vertical solid line mark $R_{\rm 0.85max}$.} 
\label{fig:bar_unbar_example}
\end{figure*}

We fit ellipses to the isodensity contours of the face-on surface density maps of the TNG100 disk galaxies with the IRAF task {\tt ellipse}. This method measures the radial profiles of ellipticity ($\varepsilon$), position angle (PA), and surface density of a galaxy. It has been widely used as an efficient method for detecting bars in observed galaxies \citep[e.g.,][]{Jogee04, MarinovaJogee07, Aguerri09, Li2011, Consolandi16}, and it can be similarly applied to our current sample of simulated galaxies. Stellar particles are binned into square bins of $350\times 350$ pc$^2$, which correspond to sides equal to half the gravitational softening radius. We use the same bar identification criteria as \citet{MarinovaJogee07}: (1) within the bar, the maximum value of $\varepsilon$ should be $> 0.25$ and the PA should vary by $< 10^\circ$; and (2) $\varepsilon$ decreases by $> 0.1$ outward from the maximum.  We refer the reader to \cite{MarinovaJogee07} for further discussion of these bar identification criteria.
In addition to automatic ellipse fits, we further visually inspect  the images to ensure reasonable morphologies of the identified barred galaxies. We find that $\sim 3$\% of the barred galaxies are misclassified as having bars, whereas they really have irregular structures in the inner regions. We exclude them to ensure no contamination in the barred galaxies. Figure~\ref{fig:bar_unbar_example} shows examples of an unbarred galaxy and two barred galaxies.

\subsection{Strength and Size of Bars}
\label{sec:measure_bar_size}

Following common practice \citep[e.g.,][]{Jogee99, Knapen00, Laine02, MarinovaJogee07, HerreraEndoqui15}, we quantify the bar strength by the maximum value of $\varepsilon$, which we designate $\varepsilon_{\rm max}$. Bars with $\varepsilon_{\rm max} \geqslant 0.4$ are classified as strong bars, and as weak bars otherwise. This threshold is the same as that used in observations \citep[e.g.,][]{Jogee04, Barazza08}.

\begin{figure}
\raggedright
\includegraphics[scale=0.7]{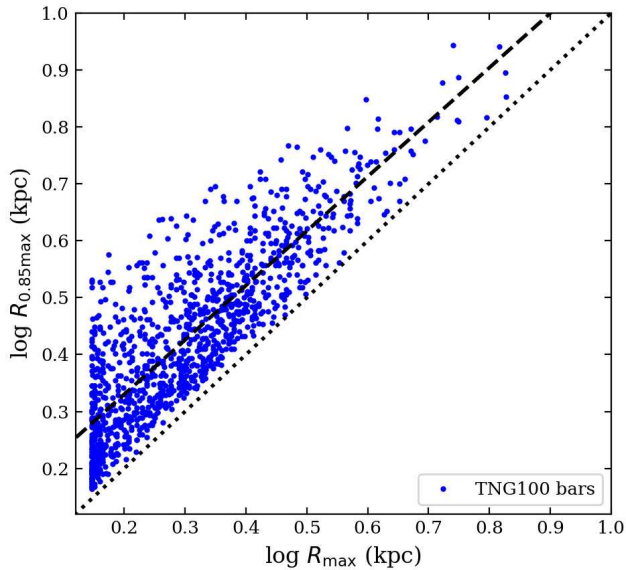} 
\caption{Comparison of $R_{\rm 0.85max}$ with $R_{\rm max}$ for TNG100 barred galaxies. The dashed line is the best-fit relation: $\log R_{\rm 0.85max}=0.96\log R_{\rm max}+0.14$. The dotted line is the 1 to 1 relation. }
\label{fig:compare_Rbar}
\end{figure}

As noted by previous studies \citep[e.g.,][]{DiazGarcia16}, bar size measurements are ambiguous and dependent on the method chosen. We adopt two methods to estimate the semi-major axis of bars to capture some of the uncertainty in bar sizes. In the first instance, $R_{\rm 0.85max}$ is the radius where $\varepsilon$ declines to 85\% of the maximal value. In their systematic study of bars based on $N$-body simulations, \citet{MartinezValpuesta2006} suggested that $R_{\rm 0.85max}$ can be used as a reliable estimator of bar size. The second choice for bar size, $R_{\rm max}$, defined as the radius at $\varepsilon_{\rm max}$, has been commonly used \citep[e.g.,][] {Wozniak1995, JungwiertCombesAxon1997, Laine2002, Sheth2003}, although both observational \citep{WozniakPierce1991, LaurikainenSalo2002, ErwinSparke2003} and theoretical \citep{RautiainenSalo1999, AthanassoulaMisiriotis02, ONeillDubinski2003, Athanassoula05, MichelDansac&Wozniak2006} studies have suggested that $R_{\rm max}$ may underestimate the true bar size. The middle and right panels of Figure~\ref{fig:bar_unbar_example} show clearly that $R_{\rm 0.85max}$ (solid ellipse) better matches the visual extent of the bar, both for the weak and strong bar examples. By comparison, $R_{\rm max}$ (dashed ellipse) somewhat underestimates the bar size.  Figure~\ref{fig:compare_Rbar} shows that $R_{\rm 0.85max}$ is statistically larger than $R_{\rm max}$ by $\sim0.14$ dex in TNG100, with the best-fit relation of $\log R_{\rm 0.85max}=0.96\log R_{\rm max}+0.14$.

An inner boundary of ellipse fitting is set at the central region of semi-major radius of 4 pixel. It excludes the regions where the results of ellipse fits are unreliable. We have visually checked that the inner boundary is neither too small nor too large to conduct reasonable bar measurements. It is worth mentioning that $R_{\rm max}$ of $\sim 10$\% of the identified barred galaxies is equal to the value of such an inner boundary of ellipse fits, due to the fact that the ellipticities of these galaxies keep increasing towards the galactic centers. We found no clear bulges formed in these galaxies, which may explain the increase of ellipticities. $R_{\rm max}$ thus is likely to underestimate bar sizes. Instead, $R_{\rm 0.85max}$ is generally consistent with our visual judgement of the bar sizes in TNG100 galaxies, exhibiting a weaker dependence on the inner boundary. Therefore, we regard $R_{\rm 0.85max}$ as the standard bar size in this paper. 

The uncertainty of $R_{\rm 0.85max}$ for each barred galaxy caused by both the artificial selection of the inner boundary and pixel size is shown by error bars in Figure~\ref{fig:bar_property}. Here we vary the inner boundary from 2.3 to 4.3 pixels. We also repeat the same analysis on the images of pixel size 175x175 pc$^2$ and 700x700 pc$^2$. These experiments show that the bar fraction and bar strength are not significantly affected by the selection of inner boundary and pixel size. Table~\ref{tab:bar} gives the bar properties for all the barred galaxies; an extract is presented here while the full table is published electronically (www.tng-project.org/zhao20).

\begin{deluxetable}{cccccc}
\fontsize{9.5}{9}\selectfont
\tablecaption{\label{tab:bar} Properties of Barred Galaxies}
\tablehead{
\colhead{Galaxy} & \colhead{$\log M_*$}          & \colhead{$k_{\rm rot}$}          & \colhead{$\varepsilon_{\rm max}$}    & \colhead{$R_{\rm max}$}    & \colhead{$R_{\rm 0.85max}$}   \\[0.5ex]
 \colhead{ID} & \colhead{($M_\odot$)}       & \colhead{  }                     & \colhead{    }                       & \colhead{(kpc)}            & \colhead{(kpc)}              \\[0.5ex]
 \colhead{(1)} & \colhead{(2)} & \colhead{(3)} & \colhead{(4)}& \colhead{(5)} & \colhead{(6)}  }

\startdata  
194950      & 10.79       & 0.58          & 0.67         & 3.40         & 4.77 $_{-0.16}^{+0.01}$         \\[0.5ex]
197110      & 10.95       & 0.57          & 0.57         & 1.80         & 3.30 $_{-0.04}^{+0.01}$         \\[0.5ex]
197112      & 10.68       & 0.60          & 0.50         & 2.45         & 3.07 $_{-0.08}^{+0.01}$         \\[0.5ex]
197114      & 10.49       & 0.63          & 0.30         & 1.40         & 1.63 $_{-0.36}^{+0.09}$         \\[0.5ex]
199322      & 10.69       & 0.62          & 0.60         & 1.41         & 2.59 $_{-0.04}^{+0.08}$         \\[0.5ex]
327406      & 10.54       & 0.55          & 0.37         & 1.71         & 2.38 $_{-0.05}^{+0.00}$         \\[0.5ex]
328253      & 10.57       & 0.56          & 0.42         & 2.35         & 3.33 $_{-0.12}^{+0.02}$         \\[0.5ex]
328679      & 10.84       & 0.58          & 0.65         & 3.51         & 5.50 $_{-0.39}^{+0.02}$         \\[0.5ex]
330209      & 10.52       & 0.57          & 0.57         & 2.47         & 3.30 $_{-0.18}^{+0.02}$         \\[0.5ex]
330956      & 10.27       & 0.53          & 0.55         & 1.47         & 1.73 $_{-0.02}^{+0.00}$         \\[0.5ex]
\enddata
\tablecomments{Col. (1): Galaxy ID. Col. (2): Stellar mass of host galaxy measured within a sphere of 30 kpc radius centered on the galaxy. Col. (3): Fraction of kinetic energy in ordered rotation. Col. (4): Bar strength, defined as the maximum ellipticity. Col. (5): Bar size measured at maximum ellipticity. Col. (6): Bar size defined as the radius where the ellipticity declines to 85\% of the maximum value, with uncertainties caused by inner boundary and pixel size. }
\label{tab:bar} 
\end{deluxetable}

\section{Bars at $z=0$}
\label{sec:result_z0}

In this section we consider the properties of bars in the TNG100 disk galaxies at $z=0$. We use surface mass density maps to derive quantities that are compared with observations, as well as to obtain the intrinsic properties of bars. 

\subsection{Bar Fractions}
\label{sec:bar_fraction_z0}

\begin{figure}
\raggedright
\includegraphics[scale=0.71]{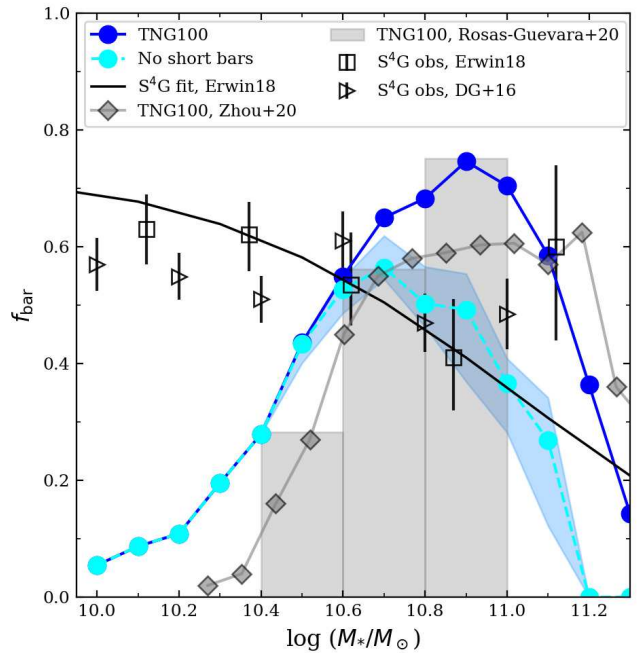}  
\caption{Bar fractions $f_{\rm bar}$ in the disk galaxies of TNG100 (filled dots). The dark blue dots represent the overall bar fraction in TNG100 while the cyan ones show the fraction if short bars in massive galaxies are excluded (as described in Section~\ref{sec:bar_size_z0}). The blue shaded area shows the uncertainty in  $f_{\rm bar}$ from variations in choosing the short bars (see Figure~\ref{fig:bar_property}). The values of $f_{\rm bar}$ from the NIR observations of S$^4$G (open symbols) are  adopted from \citet{DiazGarcia16} and \citet{Erwin18}. Bars in \citet{DiazGarcia16} are identified by ellipse fitting, as we do here. \citet{Erwin18} excludes S0 galaxies and identified bars by visual inspection, which leads to slight differences with \citet{DiazGarcia16}. The step-filled histogram and diamonds show the bar fractions estimated by \citet{Yetli20} and \citet{Zhou20}, respectively, where bars are identified by the Fourier method. }
\label{fig:comparison_obs}
\end{figure}

The fraction of barred galaxies as a function of $M_*$ in the parent disk sample is shown in Figure~\ref{fig:comparison_obs}. It is well known that bars are generally more prominent in NIR bands because of lower dust attenuation in their central regions \citep{Thronson89, BlockWainscoat91, Spillar92}. Thus, we compare the bar fraction in TNG100 with statistics derived from the S$^4$G NIR survey.  The bar fraction in TNG100 increases with stellar mass over the mass range $M_* = 10^{10}-10^{11}\, M_\odot$, while it remains almost constant, at $f_{\rm bar} \approx 0.5-0.6$, over the same mass range for the S$^4$G galaxies.  TNG100 and S$^4$G reach a similar bar fraction in galaxies with $M_* \approx 10^{10.6}\, M_\odot$. However, the discrepancy in 
$f_{\rm bar}$ becomes significant in galaxies of lower and higher mass.  Evidently TNG100 galaxies at $M_* < 10^{10.6}\, M_\odot$ host fewer bars compared to real galaxies, implying that they either do not generate them in the first place or that they have trouble maintaining them after formation.  Conversely at the massive end ($M_* \gtrsim 10^{10.6}\, M_\odot$), TNG100 produces $\sim 20$\% more bars than the NIR observations, resulting in a bar fraction up to 75\% compared with the lower S$^4$G fraction of $\sim 50$\%.  

As shown in Figure~\ref{fig:comparison_obs}, our results are roughly consistent with the trends obtained by the Fourier method in both \citet{Yetli20} and \citet{Zhou20}. Our bar fractions, however, are generally $\sim 0.1-0.2$ larger than that of \citet{Zhou20} who used a disk galaxy sample similar to ours. The reason is that the Fourier method is not as sensitive as ellipse fitting for finding short bars (see Section~\ref{sec:effects} for details).

Since TNG100 barred galaxies are identified by mass density maps, the main conclusions are based on comparison with NIR observations, which are much more reliable in determining the overall mass distribution in galaxies and minimize the possibility of missing bars due to dust obscuration or confusion from bright young stellar populations suffered in optical observations. Nevertheless, for completeness we present a comparison with SDSS-based bar fractions in Appendix \ref{app:comp_SDSS}.

Simulations of isolated disk galaxies show that bars grow more slowly in dynamically hotter disks \citep{Athanassoula&Sellwood1986}. Cosmological simulations typically suffer from overheating, particularly in less massive galaxies, whose low particle number may suppress or delay bar formation. Comparison with higher resolution runs will be needed to ascertain whether the suppressed bar fraction in the less massive galaxies is predominantely due to the low numerical resolution of TNG100.

The overproduction of bars in the more massive galaxies clearly requires a different explanation. The bar fraction in TNG100 peaks at $M_*\approx 10^{10.9}\,M_\odot$, where it reaches 75\%, while only $\sim 50\%$ of S$^4$G disk galaxies host bars. It is worth mentioning that S0 galaxies, which are highly incomplete in S$^4$G, are excluded by \citet{Erwin18} but not by \citet{DiazGarcia16}. This has a small effect on the bar fraction in massive galaxies, as illustrated in Figure~\ref{fig:comparison_obs}. In the next subsection, we study the bar sizes to explore whether the overproduction of bars in TNG100 is somehow related to their size.

\subsection{Bar Properties}
\label{sec:bar_size_z0}

\begin{figure}
\raggedright
\includegraphics[scale=0.66]{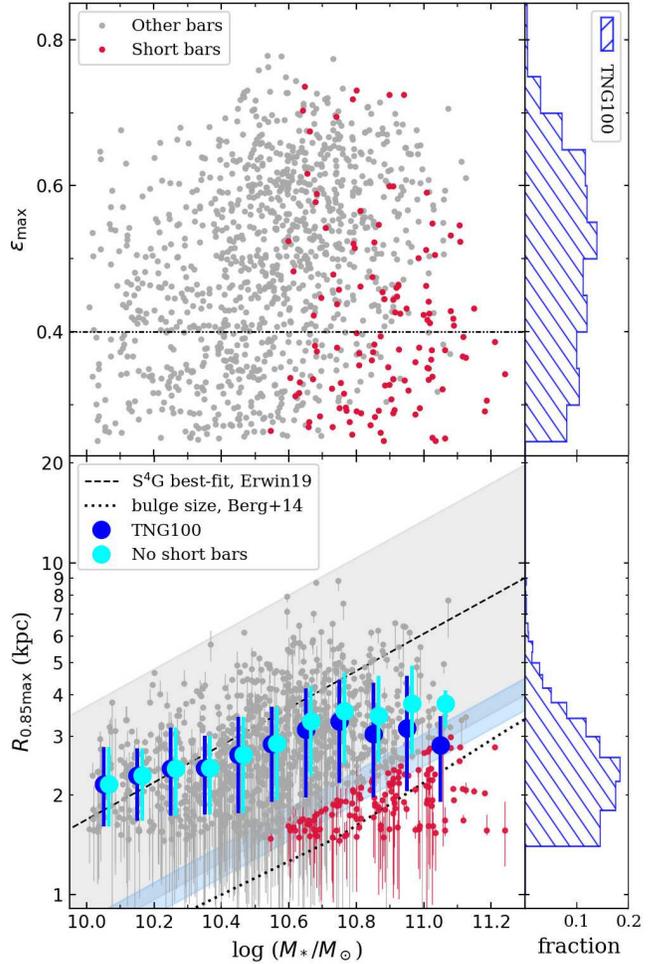}  
\caption{Distribution of (top) bar strength ($\varepsilon_{\rm max}$) and (bottom) bar size ($R_{\rm 0.85max}$) versus galaxy stellar mass ($M_*$).  In the top panel, the dash-dotted horizontal line shows the threshold used to separate strong and weak bars.  In the bottom panel, the dashed line represents the best-fit relation between bar size and stellar mass derived for S$^4$G by \citet{Erwin19}, with the grey shaded area showing the region occupied by S$^4$G barred galaxies.  The blue shaded region represents the uncertainty of the lower boundary, from the difference between the samples of \citet{DiazGarcia16} and of \citet{Erwin18,Erwin19}. The red dots highlight the massive TNG100 disk galaxies with short bars that are outside of the grey shaded area. The grey dots are the rest of the TNG100 barred galaxies. The error bars of red dots and grey dots are the uncertainties of inner boundary and pixel size. The large dark blue dots with error bars mark the median bar size and the 16 and 84 percentile at each mass bin for the overall sample of TNG100 barred galaxies; the large cyan dots give the statistics with the short bars in massive galaxies excluded. The dotted line shows the relation between stellar mass and median bulge effective radius of local galaxies \citep{Berg14}. }
\label{fig:bar_property}
\end{figure}

\begin{figure}
\raggedright
\includegraphics[scale=0.57]{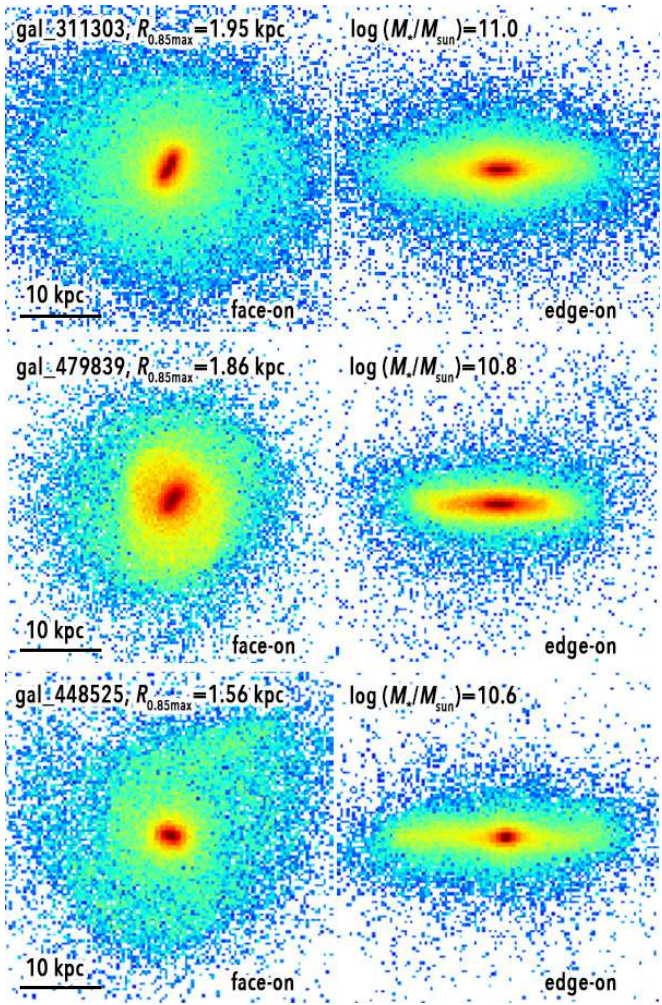}  
\caption{Examples of massive disk galaxies with relatively short bars. Their stellar masses cover $M_* = 10^{10.6}-10^{11.0}\, M_\odot$.}
\label{fig:massive_barred_gal}
\end{figure}

The top panel of Figure~\ref{fig:bar_property} shows that bar strength in TNG100 galaxies varies widely in the range $\varepsilon_{\rm max} \approx 0.2-0.8$, independent of stellar masses. This is qualititively consistent with the distribution of bar strength from NIR observations \citep[e.g.,][]{DiazGarcia16}.  

We notice in Section~\ref{sec:measure_bar_size} that $R_{\rm 0.85max}$ is consistent with the visual radius of TNG100 bar, while $R_{\rm max}$ underestimates bar size. Therefore, $R_{\rm 0.85max}$ is taken as the standard bar size in this paper to compare with the S$^4$G results of \cite{Erwin18, Erwin19}. As shown in \cite{Erwin18, Erwin19}, many bars in S$^4$G galaxies with $M_*\lesssim 10^{10.6}\, M_\odot$ are shorter than 1.4 kpc. Because we run the ellipse fitting with a inner boundary, the lower limit of the measured $R_{\rm 0.85max}$ is close to 1.4 kpc. Thus, galaxies with bar sizes smaller than 1.4 kpc cannot be identified in TNG100, partially explaining the drop in the bar fraction in the mass range of $M_* \lesssim 10^{10.6}\, M_\odot$ (Figure~\ref{fig:comparison_obs}).

\cite{DiazGarcia16} and \cite{Erwin18, Erwin19} showed that bars follow an empirical relation between bar size and total stellar mass, becoming larger in galaxies with $M_* \gtrsim 10^{10}\, M_\odot$. TNG100 barred galaxies follow a similar, but shallower, trend in the mass range $M_* \approx 10^{10.2}-10^{10.8}\, M_\odot$ (large dark blue dots with error bars in the bottom panel of Figure~\ref{fig:bar_property}). In this mass range, the median size of the TNG100 bars is slightly shorter than that measured in the S$^4$G survey. Relatively shorter bars (red dots) are present toward the high-mass end, falling below the lower boundary of \citet[][grey shaded region\footnote{The grey shaded region in Figure~\ref{fig:bar_property} is defined by requiring that no S$^4$G barred galaxies lie outside this region.}]{Erwin19}. Three examples of such short bars are shown in Figure~\ref{fig:massive_barred_gal}.

The question of whether S$^4$G can detect short bars in the massive galaxies therefore naturally arises. Using the S$^4$G spiral galaxies at distance of $\lesssim 30$ Mpc, \citet{Erwin18} claimed that bars as short as $0.33$ kpc can be resolved, as confirmed by the high incidence of short bars in less massive galaxies ($M_*< 10^{10}\, M_\odot$; \citealp{Erwin18}). A possible observational bias can arise if short bars are missed when they coexist with a massive bulge, which lowers the contrast and lowers $\varepsilon_{\rm bar}$. The black dotted line in the bottom panel of Figure~\ref{fig:bar_property} represents the median effective radii of bulges derived from over $10^5$ local galaxies \citep{Berg14}. Bars with sizes longer than the bulge effective radii are unlikely to be missed in observations. Accordingly, the lack of massive galaxies with relatively short bars detected by S$^4$G, especially the absence of barred galaxies in the region between the lower boundary of the grey shaded region and the black dotted line, implies that such short-bar galaxies are extremely rare in the local Universe.  The foregoing considerations leads us to conclude that TNG100 overproduces short bars in massive galaxies.  If these excess relatively short bars (red dots in Figure~\ref{fig:bar_property}) are counted as unbarred galaxies, the overall bar fraction (cyan dots in Figure~\ref{fig:comparison_obs}) come into a good agreement with the NIR observations. Though the median values of bar sizes (large cyan dots in Figure~\ref{fig:bar_property}) are still somewhat smaller than the observations, the discrepancy is not significant, considering the large uncertainty in bar measurements.

The bar sizes used in \cite{Erwin18, Erwin19} were estimated visually. Their results are comparable with those obtained by $R_{\rm 0.85max}$ in TNG100. As $R_{\rm max}$ is smaller than $R_{\rm 0.85max}$, there is no doubt that an even larger number of short bars will be found in comparison with the bar sizes from S$^4$G. If these short bars in massive galaxies are counted as unbarred galaxies, the overall bar fraction and bar sizes become more consistent with the NIR observations. Therefore, different definitions of bar size do not change our main conclusions.

Moreover, in \citet{Yetli20} bar sizes are defined as the radius at the maximum $A_2$. \cite{DiazGarcia16} showed that visually estimated bar sizes are larger than those defined at $A_2$ maximum by a factor of $\sim 1.3$. Therefore, it is not surprising that the bar sizes measured by \citet{Yetli20} could be as small as $\sim 1.1$ kpc. Consistently, Figure 3 of \citet{Yetli20} also shows some short bars in massive galaxies which are scarcely observed in observations.

\subsection{Uncertainties Due to Selection Criteria}
\label{sec:effects}

The bar fraction may also be affected by the criteria used to select disk galaxies in the first place. Can the difference between TNG100 and observations be caused by the selection criteria for disk galaxies? We investigate this issue here.  

The literature employs five widely used sets of criteria to select disk galaxies:

\begin{enumerate}
    \item Konly: All rotation-dominated galaxies are included by setting the rotation criterion $k_{\rm rot} \geqslant 0.5$. This is the criterion used in this paper. 
 
    \item KF: In addition to $k_{\rm rot} \geqslant 0.5$, the morphological flatness criterion $M_1/\sqrt{M_2 M_3} \leqslant 0.5$ is imposed, where where $M_1$, $M_2$, and $M_3$ are the eigenvalues of the mass tensor of the stellar mass inside $2r_e$ and $M_1<M_2<M_3$.  These criteria select rotation-dominated galaxies with disky morphologies and are similar to those employed by \citet{Algorry17} in the EAGLE simulation.  
 
    \item KFsSFR: This set of criteria selects disk galaxies satisfying not only the KF criteria but also require the specific star formation rate ${\rm sSFR} > 10^{-10.75}\, \rm{yr}^{-1}$, defined as the ratio between the star formation rate (SFR) and stellar mass. The SFR is derived by summing up the stars formed over the last 1 Gyr within $2r_e$ \citep{Donnari19}. Disk galaxies selected by these criteria are likely to be mainly late-type galaxies on the star-forming main sequence \citep[see][]{Donnari19, Xu19}, avoiding early-type galaxies. 
     
    \item DTF: Disk galaxies are selected by requiring a disk-to-total mass ratio $D/T\geqslant 0.2$ and flatness $M_1/\sqrt{M_2 M_3} \leqslant 0.7$. The mass fraction of the disk component is kinematically derived by the method described in \cite{Abadi03}, \cite{Marinacci2014}, and \cite{Genel15}, which sums up the stellar particles within $2r_e$ having circularity parameter $\epsilon > 0.7$. This criterion has been applied in both the original Illustris \citep{Peschken19} and TNG100 \citep{Zhou20} data.
 
    \item BDT: Disk galaxies are selected by  $D/T\geqslant 0.5$ and $(B/T + D/T)\geqslant 0.7$, where the kinematically derived mass fraction of the bulge ($B/T$) is twice the mass fraction of counter-rotating stellar particles ($\epsilon<0$). These criteria, used by \citet{Yetli20} for TNG100, tend to select strongly rotating disk galaxies, while galaxies that host significant warm components (e.g., thick disks and pseudo bulges) are excluded.     

\end{enumerate}

\begin{figure}
\raggedright
\includegraphics[scale=0.72]{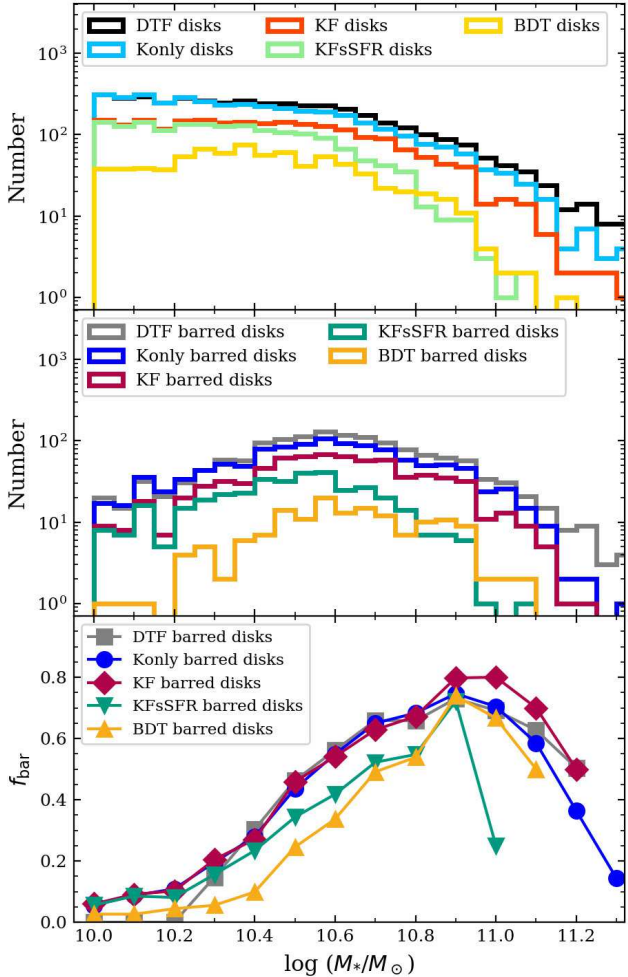}  
\caption{Effect of the criteria used to select disk galaxies on the resulting bar statistics. The disk galaxies are selected by five sets of criteria: Konly, KF, KFsSFR, DTF, and BDT. The top, middle and bottom panel show the distributions as a function of stellar mass of the disk galaxies and of the barred galaxies, and of the resulting bar fractions, respectively.}
\label{fig:diff_disk_number}
\end{figure}

\begin{figure}
\raggedright
\includegraphics[scale=0.57]{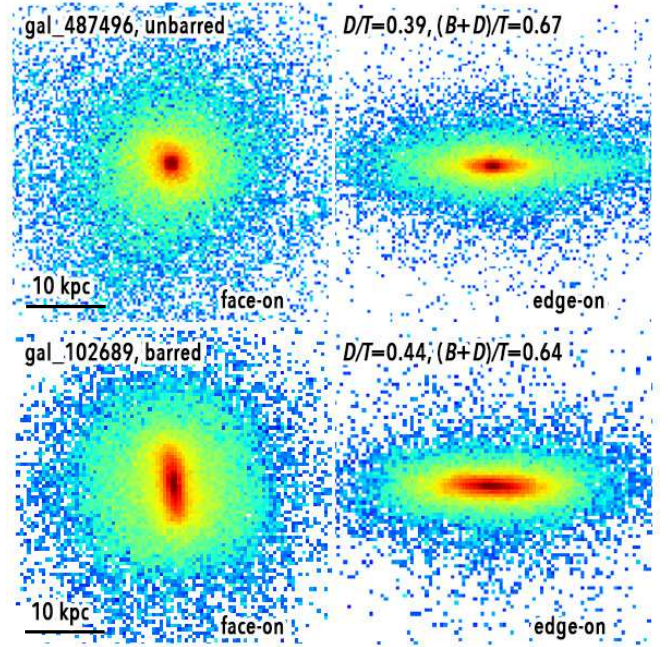}  
\caption{Examples of an unbarred (top panels) and a barred (bottom panels) galaxy, viewed face-on and edge-on.  Both galaxies are selected as disks ccording to the Konly criterion,  but not using the BDT criterion. They both clearly have a disk morphology, and the bottom galaxy exhibits quite a long, obvious bar.}
\label{fig:undisk_BDT_sample}
\end{figure}

The five sets of criteria (Konly, KF, KFsSFR, DTF, and BDT) select 3866, 2291, 1664, 4015, and 795 disk galaxies, yielding 1182, 689, 299, 1316, and 163 barred galaxies, respectively.  Apparently, the selection criteria become progressively stricter from Konly to KF to KFsSFR, resulting in fewer total disk and barred galaxies selected. The BDT criteria give the lowest number of galaxies. 

The top and middle panel of Figure~\ref{fig:diff_disk_number} show that KF (red and purple) gives a similar sample of disk (barred) galaxies as Konly (cyan and blue) in the high-mass end, but relatively fewer galaxies at low masses, suggesting that many rotation-dominated low-mass galaxies appear too thick and relatively spheroidal. Adding the sSFR condition (KFsSFR; lime and green) further removes relatively more quiescent galaxies at the high-mass end; this would select against S0 galaxies and bias the sample toward late-type disk galaxies.

The DTF criterion (black and grey) selects a similar sample of disk (barred) galaxies as our adopted standard Konly criterion. However, our ellipse fitting method identifies $\sim 10\%$ more barred galaxies compared to the Fourier method applied by \citet{Zhou20} (see Figure~\ref{fig:comparison_obs}). Comparing the barred galaxies identified independently by these two methods, we verified that the Fourier method is not as sensitive as ellipse fitting for finding short (bar size less than $2.5$ kpc) and weak (bar strength less than 0.4) bars. Within the mass range in which our studies overlap ($M_* \geqslant 10^{10.25}\, M_\odot$), only $22\%$ of the short and weak bars we found are identified successfully in \citet{Zhou20}. For stronger and longer bars, the difference between these two methods are negligible. It is worth emphasizing that 43\% of the short bars overproduced in TNG100 massive ($M_* \geqslant 10^{10.6}\, M_\odot$) galaxies (see Section \ref{sec:bar_size_z0}) are also identified by \citet{Zhou20}.  Thus, our conclusion that TNG100 overproduces short bars in massive galaxies is confirmed qualitatively in the analysis of \citet{Zhou20}.

The BDT criteria (yellow and gold) select the smallest sample over the entire mass range, missing a large number of disk galaxies, barred or otherwise. Applying these criteria, \citet{Yetli20} select only 270 disk galaxies with $M_* \gtrsim 10^{10.4}\, M_\odot$. Figure~\ref{fig:undisk_BDT_sample} shows an example of an unbarred and barred galaxy selected by Konly but excluded from the BDT sample. There is little doubt that these cases should have been included.  There are two possible reasons why the BDT criteria miss the majority of disk galaxies in TNG100, especially those with strong (bar strength larger than $0.55$) and long (bar size larger than $3$ kpc) bars. First, many TNG100 galaxies have a massive warm disk. As shown by \citet[][Figure 9]{Du2020}, the kinematically derived warm disk contributes $\sim 35\%$ of the total stellar mass. Secondy, particles on bar orbits necessarily have smaller circularity, and hence they are more likely to be classified as belonging to the hotter components instead of the cold disk\footnote{See Figure 12 in \citet{Du2019} for an example.}. These reasons probably explain why we find that BDT misses disk galaxies with strong (bar strength larger than $0.55$) and long (bar size larger than $3$ kpc) bars.

Although different criteria select very different samples of disk galaxies, their bar fractions follow a surprisingly similar tendency to increase systematically with stellar mass (bottom panel of Figure~\ref{fig:diff_disk_number}). The uncertainty in bar fraction resulting from differences in selection criteria is less than 15\% across almost the entire mass range. Thus, the inconsistency between TNG100 and observations cannot be explained by the criteria used to select disks in TNG100.

\section{Evolution of Bars}
\label{sec:result_zgt0}

\begin{figure}
\raggedright
\includegraphics[scale=0.74]{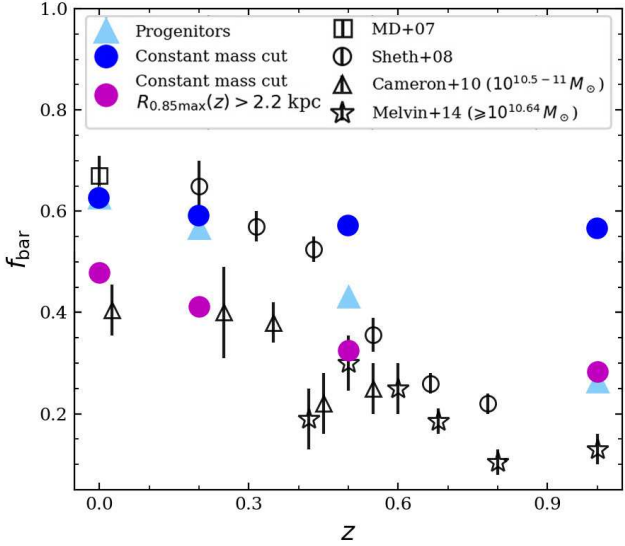}  
\caption{Bar fraction ($f_{\rm bar}$) as a function of redshift. Solid blue dots correspond to the bar fraction evolution for the sample selected by a constant mass cut of $M_* \geqslant 10^{10.6}\, M_\odot$. Solid magenta dots show the bar fraction evolution with an additional criterion of $R_{\rm 0.85max}(z)>2.2$ kpc. The solid light blue triangles show bar fraction evolution obtained by tracing the progenitors of the galaxy sample selected at $z=0$. Observational data are from \citet{MDelmestre07}, \citet{Sheth08}, \citet{Cameron10}, \citet{Melvin14}. }
\label{fig:barfraction_evolve}
\end{figure}

\begin{figure*}
\centering
\includegraphics[scale=0.68]{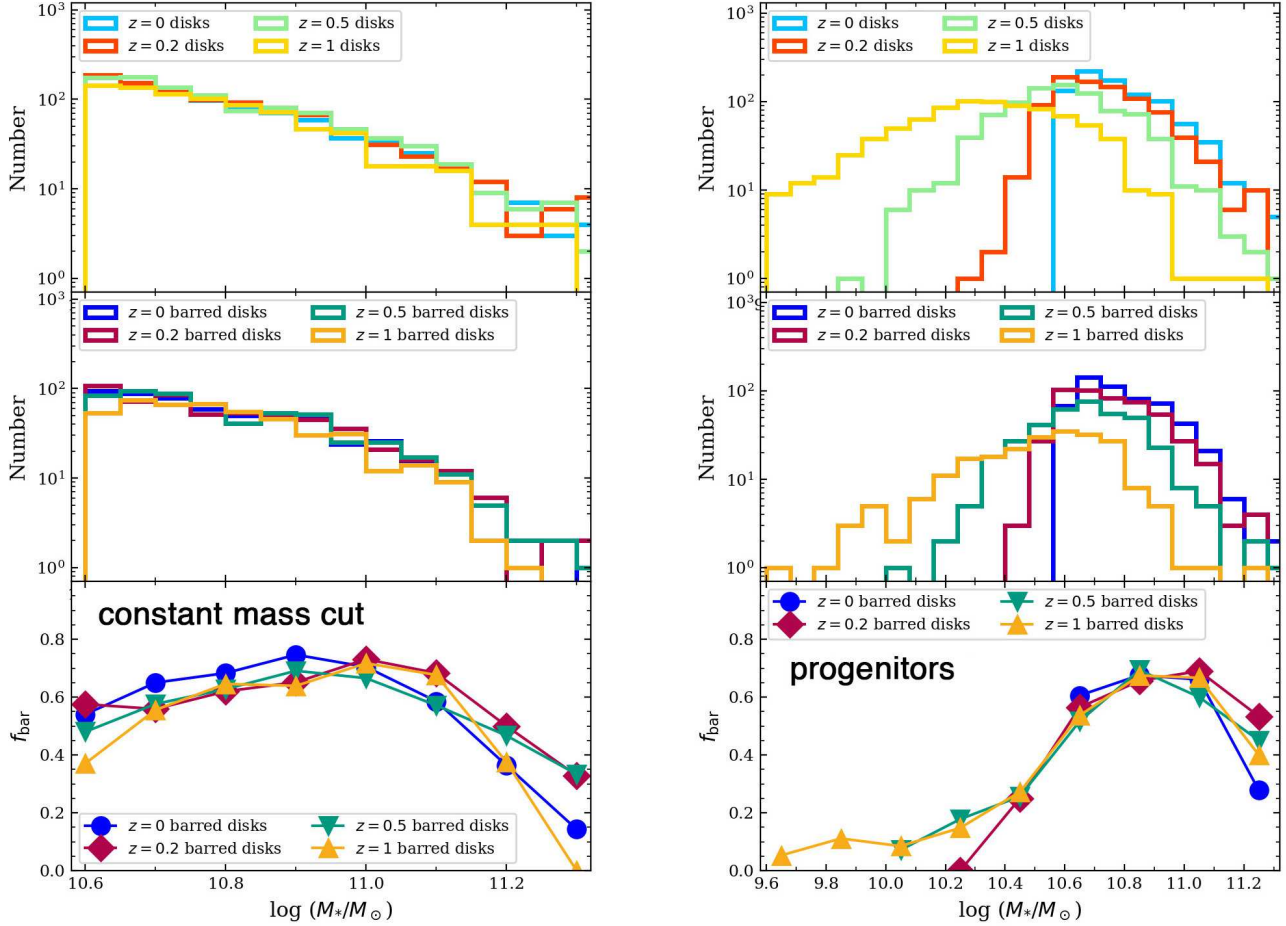}  
\caption{From top to bottom: the number distributions of disk and barred galaxies, and the bar fraction binned by stellar mass. The results measured at redshifts $z=0, 0.2, 0.5$ and 1.0 are shown. The left column shows the results of the sample of disk galaxies selected by a constant mass cut of $M_* \geqslant 10^{10.6}\, M_\odot$, and the right column is the results obtained by tracing the progenitors of the $z=0$ sample.}
\label{fig:number_fraction_evolve}
\end{figure*}

\begin{figure}
\raggedright
\includegraphics[scale=0.58]{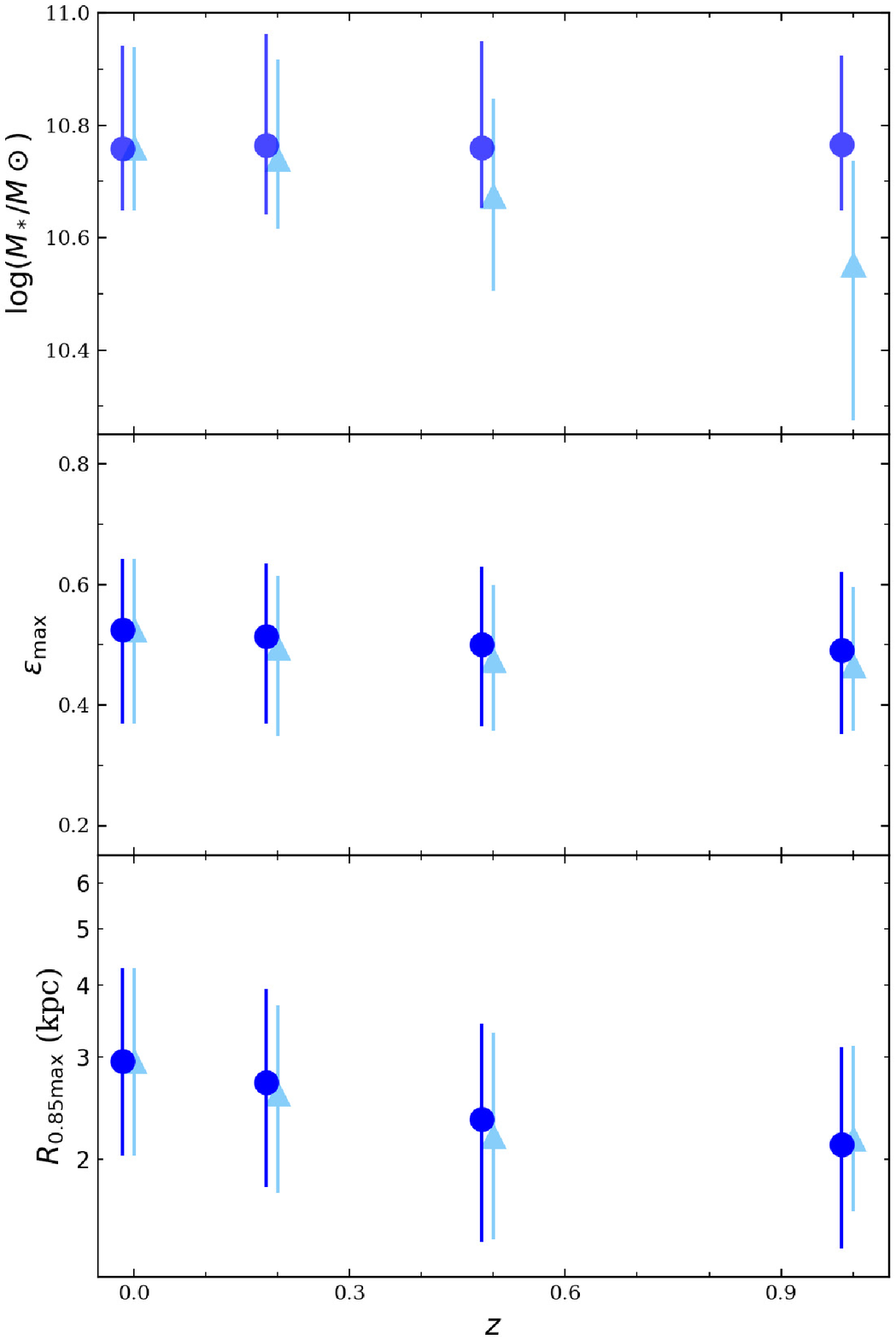}  
\caption{Cosmic evolution of stellar mass (top), bar strength ($\varepsilon_{\rm max}$; middle) and bar size ($R_{\rm 0.85max}$; bottom) of TNG100 barred galaxies. The blue dots are the results with constant mass cut of $M_* \geqslant 10^{10.6}\, M_\odot$, and light blue triangles are the results by tracing the  progenitors.
Each point shows the median value at a given redshift, with error bars corresponding to the 16 and 84 percentiles.}
\label{fig:bar_evolve_prop}
\end{figure}

We study the evolution of the bar fraction at $z=0$ to $1$ using two methods. First, at all redshifts, the sample of massive disk galaxies is selected using the same criteria, $k_{\rm rot} (z) \geqslant 0.5$ and a constant mass cut of $M_* \geqslant 10^{10.6}\, M_\odot$. The ellipse fitting method is applied to the face-on view of the mass surface density maps. Figure~\ref{fig:barfraction_evolve} shows that the total bar fraction (solid blue dots) is nearly constant, increasing slightly from 0.57 at $z=1$ to 0.63 at $z=0$. The left column of Figure~\ref{fig:number_fraction_evolve} shows that, using this method, disk or barred galaxies have a similar number distribution over $z=0-1$, and the bar fraction at all redshifts roughly follows a similar trend. Additionally, Figure~\ref{fig:bar_evolve_prop} shows that during the past 6 Gyr bars have maintained nearly constant strength ($\varepsilon_{\rm max} \approx 0.5$), while the bar sizes have grown by 0.17 dex.

The second method aims to trace the progenitors of $z=0$ disk galaxies with $M_* \geqslant 10^{10.6}\, M_\odot$. The  progenitor of each galaxy corresponds to the most massive one in its merger tree created by the {\tt SUBLINK} algorithm \citep{RodriguezGomez2015}. We have confirmed that most of the  progenitors are already rotation-dominated objects with $M_*\geqslant 10^{10}\, M_\odot$ since $z=1$. Figure~\ref{fig:barfraction_evolve} shows that the total bar fraction (solid light blue triangles) increases quickly from $f_{\rm bar} = 25\%$ at $z = 1$ to $f_{\rm bar} = 63\%$ at $z=0$. This result suggests that a large fraction of bar structures form at $z<1$. As shown in the right column of Figure~\ref{fig:number_fraction_evolve}, most bar structures form when disk galaxies grow more massive than $M_* \gtrsim 10^{10.5}\, M_\odot$. Consistent with the results of the first method, the bar strengths are nearly constant, and the bar sizes grow similarly during $z=0-1$. The increase of bar size with time qualitatively supports the picture that bars grow longer from outward transport of angular momentum \citep{Debattista&Sellwood2000, Athanassoula2004}. The results discussed above suggest that bars have already existed in a large fraction of massive disk galaxies at $z=1$. However, many such high-$z$ massive (barred) disk galaxies become elliptical galaxies with $k_{\rm rot} (z)<0.5$, possibly due to mergers.

In Figure~\ref{fig:barfraction_evolve}, we also compare the evolution of bar fractions between observations and the simulation. It is worth emphasizing that it is very difficult to make an accurate comparison between simulation results and observations at high redshifts, mainly due to the large uncertainty in the sample selection and bar measurement. Thus, here we only make a qualitative comparison. The high resolution of the \textit{Hubble Space Telescope (HST)} enables the detection of bar structures at high redshifts. Using a large sample of spiral galaxies in the COSMOS field, \citet{Sheth08} showed that the bar fraction increases from $\sim 20\%$ at $z\sim 0.8$ to $\sim 65\%$ at $z\sim 0.2$ by tracing back all galaxies brighter than $L_V^*$ with an empirically determined luminosity evolution in \cite{Capak2003} ($M_V^*=-21.7$ mag at $z=0.9$). Additionally, \cite{Cameron10} demonstrated an increase bar fraction during $z=0.6-0.2$ for COSMOS disk galaxies with $10^{11}\, M_\odot \geqslant M_*\geqslant 10^{10.5}\, M_\odot$. \cite{Melvin14} also suggested a sharp increase of overall bar fraction during $z=1.0-0.4$ for massive disk galaxies with $M_*\geqslant 10^{10.64}\, M_\odot$ selected from the COSMOS sample.

The methods used in these observations to trace the cosmic evolution of bars are similar to our first method using a constant mass cut. There is a clear discrepancy between TNG100 bar fractions (blue dots) and observations (open points), as shown in Figure~\ref{fig:barfraction_evolve}. It may be due to the fact that only bars with size $\gtrsim 2$ kpc can be detected at $z\sim 1$ by the \textit{HST} images, while high-$z$ bars measured in TNG100 can be as small as $1.4$ kpc. If only bars of radius $>2.2$ kpc are taken into account at all redshifts to select barred galaxies, which is the same as \cite{Cameron10}, the bar fraction evolution (solid magenta dots in Figure~\ref{fig:barfraction_evolve}) becomes roughly consistent with the results from \cite{Cameron10} (open triangles). Therefore, the discrepancy of bar fractions between TNG100 and observations may be due to the failure of observations to detect short bars. Moreover, short bars overproduced in TNG100 massive galaxies, as presented in Section~\ref{sec:bar_size_z0}, may also contribute to the discrepancy between the simulation and observations.

\section{Conclusions}
\label{sec:conclude}

We systematically study the properties of barred galaxies in the cosmological simulation \TNG. Bars are identified from ellipse fitting of the face-on mass surface density maps of 3866 disk galaxies selected from a parent sample of 6507  $z = 0$ galaxies with stellar masses $M_* \geq 10^{10.0}\,M_\odot$ produced from the TNG100 run. A detailed catalog, including the strengths and sizes of the bars, is publicly released with this paper. 

TNG100 represents remarkable progress in that a significant fraction of disk galaxies generate reasonable bars. About 55\% of disk galaxies with stellar mass $\sim 10^{10.6}\, M_\odot$ host bars, in agreement with observations. The bars in the simulated galaxies follow a bar size-stellar mass scaling relation that is roughly consistent with NIR observations. Notwithstanding these significant successes, some clear discrepancies with observations remain. For more massive galaxies ($M_* > 10^{10.6}\, M_\odot$), the TNG100 bar fraction is $\sim 10\%-20\%$ higher than observed in the NIR.  We attribute this to an excess population of short (radius $\sim 1.4-3$ kpc) bars in the simulations.  The predicted bar fractions align better with observations if the massive galaxies with short bars are counted as unbarred galaxies, suggesting that TNG100 overproduces, or retains, too many short bars in massive disk galaxies.  At the same time, the bar fraction of galaxies with $M_* < 10^{10.6}\, M_\odot$ decreases drastically toward the low-mass end, in sharp conflict with observations.  We attribute this discrepancy to the inability of the present modest resolution of TNG100 to detect bars with radii $\lesssim 1.4$ kpc. 

Two methods are applied to trace the evolution of TNG100 bar fraction during $z=0-1$. One is to select disk galaxies at each snapshot using a constant mass cut of $M_* \geqslant 10^{10.6}\, M_\odot$. The other one is to trace the progenitors of $z=0$ massive disk galaxies of $M_* \geqslant 10^{10.6}\, M_\odot$. The bar fraction in TNG100 disk galaxies are nearly constant at $\sim 0.6$ by constant mass cut, while it decreases dramatically in observations using a similar sample selection. We suggest that observations may fail in identifying short bars of radius $\lesssim 2$ kpc at high redshifts, or TNG100 have produced too many bars since $z=1$.

\acknowledgments
This work was supported by the National Science Foundation of China (LCH: 11721303, 11991052), the National Key R\&D Program of China (LCH: 2016YFA0400702), the Scholar Program of Beijing Academy of Science and Technology (DZ: BS202002), the China Postdoctoral Science Foundation (DZ: 2017M620499; MD: 8201400927), the National Postdoctoral Program for Innovative Talents (MD: 8201400810), and Science and Technology Facilities Council Consolidated (VPD: ST/R000786/1). The TNG100 simulation used in this work, one of the flagship runs of the IllustrisTNG project, was run on the HazelHen Cray XC40-system at the High Performance Computing Center Stuttgart as part of project GCS-ILLU of the Gauss Centres for Supercomputing. The authors thank all the members of the IllustrisTNG team for making the simulation data available to us prior to public release. Our analysis used the High-performance Computing Platform of Peking University.

\appendix

\section{Comparison with SDSS Observations}
\label{app:comp_SDSS}

\citet{RodriguezGomez19} generate mock SDSS images for TNG100 galaxies using the {\tt SKIRT} radiative transfer code, taking into consideration the effects of the point-spread function and dust attenuation and scattering. The bar fractions estimated using the mock SDSS $r$-band images can be directly compared with those derived from SDSS observations. Since the mock images of TNG100 galaxies are randomly oriented, only the disk galaxies with inclination angle $i\leqslant 60^\circ$ are selected\footnote{Disk inclination is derived from $\cos i =1-\varepsilon_{\rm disk}$, where $\varepsilon_{\rm disk}$ is the ellipticity of the disk at $2r_e$.}. This is satisfied by 2790 out of the 3866 disks galaxies initially selected by the Konly criterion. To ensure proper comparison with the corresponding SDSS observations, barred galaxies are identified both by ellipse fitting and by visual inspection.

\begin{figure*}[htbp]
\centering
\includegraphics[scale=0.8]{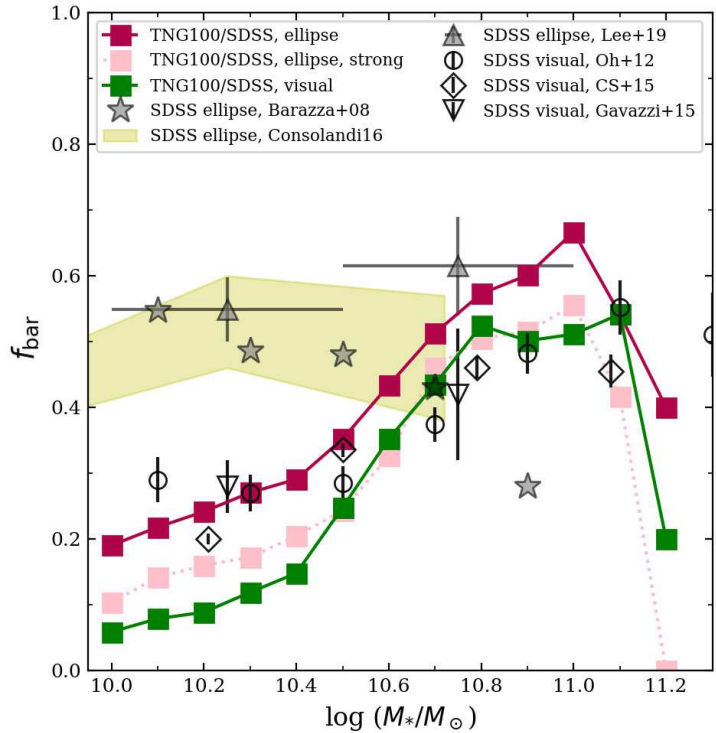}  
\caption{Bar fractions ($f_{\rm bar}$) in TNG100 using mock SDSS images. Dark red squares show $f_{\rm bar}$ determined from ellipse fitting, with the statistics for strong bars highlighted as pink squares. Dark green squares show the $f_{\rm bar}$ determined by visual inspection. For comparison, we show the SDSS-based values of $f_{\rm bar}$ derived from ellipse fitting \citep{Barazza08, Consolandi16, Lee19} and from visual inspection \citep{Oh12, Sodi15, Gavazzi15}.}
\label{fig:comparison_obs_SDSS}
\end{figure*}

Figure~\ref{fig:comparison_obs_SDSS} shows that the bar fraction increases as a function of stellar mass, with the dark red squares showing the ellipse fitting results and the dark green squares showing those based on visual inspection. The bar fractions based on ellipse fitting are larger than those based on visual classification, in agreement with observational studies \citep[e.g.,][]{Oh12, Lee19}, which find that ellipse fitting is more sensitive to weak bars.  We reconfirm this by showing the fraction of strongly barred TNG100 galaxies identified by ellipse fitting (pink squares), which is consistent with the visually estimated  bar fraction (dark green squares). Moreover, we find that bars identified in mock images are larger than 2.5 kpc in radius, which implies that the short bars detected in massive galaxies (see Section~\ref{sec:bar_size_z0}) cannot be resolved by mock images. \citet{Erwin18} previously noted that bar radius smaller than $\sim 2.5$ kpc cannot be detected in SDSS images.

We first compare $f_{\rm bar}$ from visual inspection (dark green squares) with SDSS-based visual bar fractions \citep[i.e.,][]{Oh12, Sodi15, Gavazzi15}. At $M_* < 10^{10.6}\, M_\odot$, TNG100 exhibits lower $f_{\rm bar}$ than SDSS, probably due to low numerical resolution. In contrast, TNG100's $f_{\rm bar}$ shows excellent agreement with SDSS for $M_* \gtrsim 10^{10.6}\, M_\odot$.  The mock SDSS images reveal as many long and strong bars as optical SDSS observations.  Turning to the results derived from ellipse fitting, TNG100 yields considerably lower $f_{\rm bar}$ than observed at $M_* < 10^{10.6}\, M_\odot$ but achieves better consistency at $M_* \approx 10^{10.6}-10^{10.7}\, M_\odot$ \citep{Barazza08, Consolandi16, Lee19}.  The dearth of SDSS data at $M_* > 10^{10.7}\, M_\odot$ precludes a meaningful comparison at the highest mass end.

Note that \citet{Barazza08} obtain a lower $f_{\rm bar}$ at $M_* \approx 10^{10.9}\, M_\odot$ compared to other studies. We suspect that this is caused by differences in morphological type.  While most studies mainly consider Sa--Sb spirals at the massive end, \citet{Barazza08} only focus on Sd and Sm spirals, whose bar fraction is about 20\% lower than that of Sa--Sb spirals \citep{Giordano10}.

\bibliographystyle{aasjournal} 
\bibliography{Bargal_TNG_paper1}

\end{document}